\documentclass[aps,prb,twocolumn,groupedaddress,showpacs]{revtex4}
\usepackage[colorlinks=true]{hyperref} 
\hypersetup{urlcolor=blue,linkcolor=red,citecolor=blue,colorlinks=true}

\usepackage{graphics,epsfig}
\usepackage{amssymb,amsfonts}
\usepackage{amsmath}
\usepackage{mathtools}
\usepackage{color}
\usepackage{pgf,tikz}

\newcommand{\nn}{\nonumber}
\newcommand{\bb}[0]{\begin{eqnarray}}
\newcommand{\ee}[0]{\end{eqnarray}}

\newcommand{\bara}[1]{\bar a_{#1}}
\newcommand{\barb}[1]{\bar b_{#1}}
\newcommand{\eref}[1]{Eq.~(\ref{#1})}
\newcommand{\efig}[1]{Fig.~\ref{#1}}
\newcommand{\Pf}{{\rm Pf}}
\newcommand{\Tr}[1]{\mathrm{Tr}_{\{#1\}} }

\newcommand{\ff}{\frac{1}{2}}
\newcommand{\ffi}{\frac{i}{2}}

\newcommand{\prodd}[2]{\overrightarrow{\prod_{#1}^{#2}}}
\newcommand{\prodg}[2]{\overleftarrow{\prod_{#1}^{#2}}}
\newcommand{\br}{{\bf r}}
\newcommand{\PQ}[2]{\mathcal{Q}_{#1}(\{#2\})}

\newcommand{\vc}[1]{{\bf c}_{#1}}
\newcommand{\bw}{\begin{widetext}}
\newcommand{\ew}{\end{widetext}}
\renewcommand{\d}{{\rm d}}
\newcommand{\<}{\langle}
\renewcommand{\>}{\rangle}
\newcommand{\e}{{\rm e}}

\newcommand{\dst}{d}
\newcommand{\da}{{\rm 1D}}
\newcommand{\daa}{{\rm 2D}}
\newcommand{\daaa}{{\rm 3D}}
\newcommand{\tc}{\textcolor{black}}
\newcommand{\Sdimer}{\mathcal{S}_0}

\begin{document}
\title{Grassmannian representation of the two-dimensional monomer-dimer 
model}

\author{Nicolas Allegra}
\email{nicolas.allegra@univ-lorraine.fr}
\author{Jean-Yves Fortin}
\email{jean-yves.fortin@univ-lorraine.fr}

\affiliation{
Institut Jean Lamour, CNRS/UMR 7198, Groupe de Physique 
Statistique, 
Universit\'e de Lorraine, BP 70239, F-54506 Vand{\oe}uvre-l\`es-Nancy 
Cedex, France}

\begin{abstract}
We present an application of the Grassmann algebra to the problem of 
the monomer-dimer statistics on a two-dimensional square lattice. 
The exact partition function, or total number of possible configurations, 
of a system of dimers with a finite set of $n$ monomers at fixed positions 
can be expressed via a quadratic fermionic theory. We give an answer in 
terms of a product of two pfaffians and the solution is closely related to the Kasteleyn result 
of the pure dimer problem. 
Correlation functions are in agreement with previous results, both for 
monomers on the boundary, where a simple exact expression is available in the 
discrete and continuous case, and in the bulk where the expression is evaluated numerically.  
\end{abstract}

\date{\today}

\maketitle

The study of the classical dimer model has a very long history in physics and 
mathematics. This model is interesting as a direct physical representation, e.g. diatomic molecules on a two-dimensional subtrate 
\cite{fowler1937attempt}. From the mathematical point of view, this model
on bipartite lattice -- known as a special case of perfect matching problem 
\cite{plummer1986matching}-- is a famous and active problem of combinatorics 
and graph theory \cite{flajolet2009analytic}.
The partition function of the $\daa$ dimer model was solved independently using 
pfaffian methods 
\cite{kasteleyn1961statistics,fisher1961statistical,temperley1961dimer}, 
resulting in the exact calculation of \tc{correlation functions of 
two monomers along a row (or a column) \cite{fisher1963statistical} or along a 
diagonal \cite{hartwig1966monomer,fisher2009toeplitz} in the infinite square 
lattice limit using Toeplitz determinants. For the general case of an arbitrary 
orientation, exact results are given in terms of the pair correlations of the 
$\daa$ square lattice Ising model at the critical point 
using recurrence relations\cite{perk84,kong87}.} 

\begin{figure}
\includegraphics[scale=0.054]{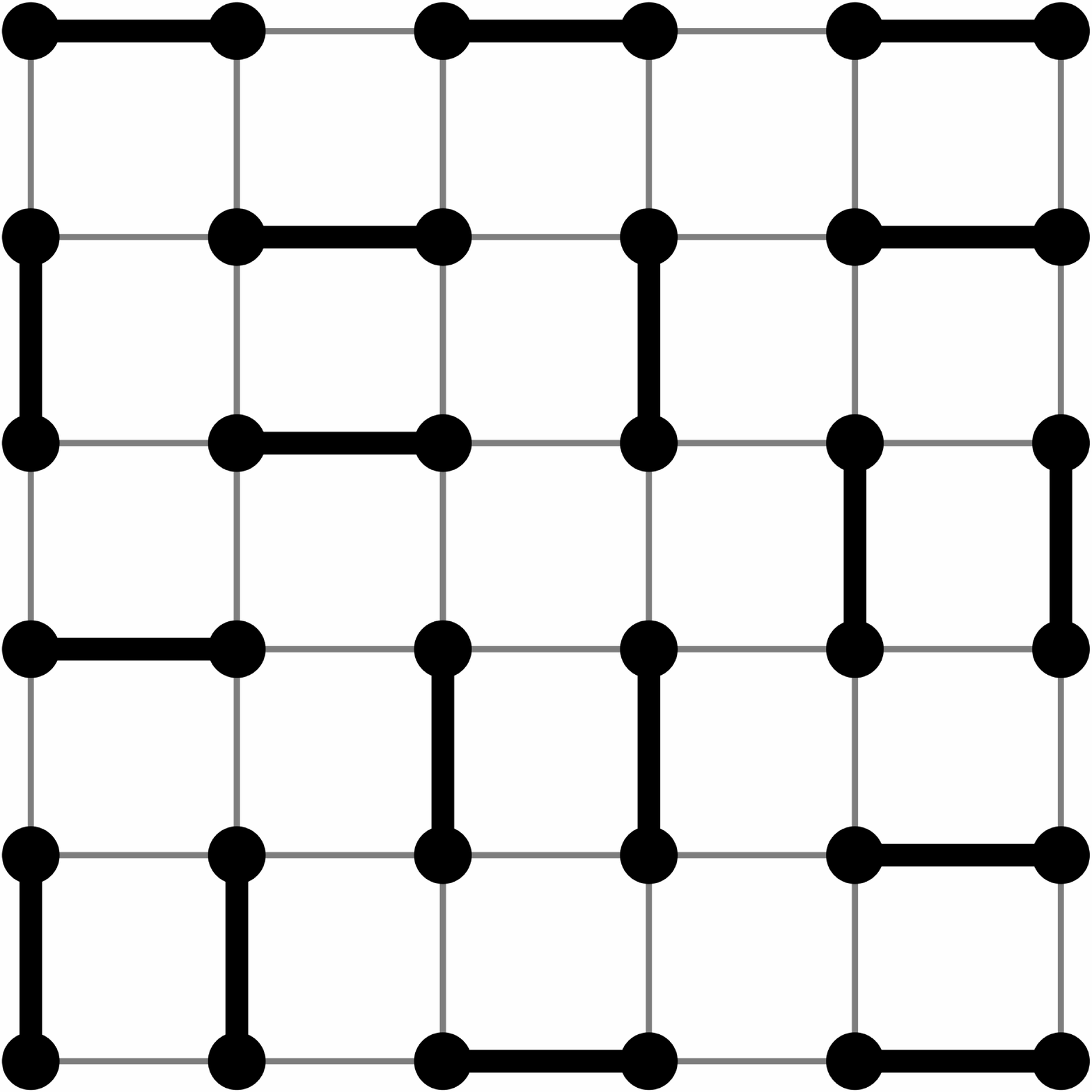} 
\includegraphics[scale=0.055]{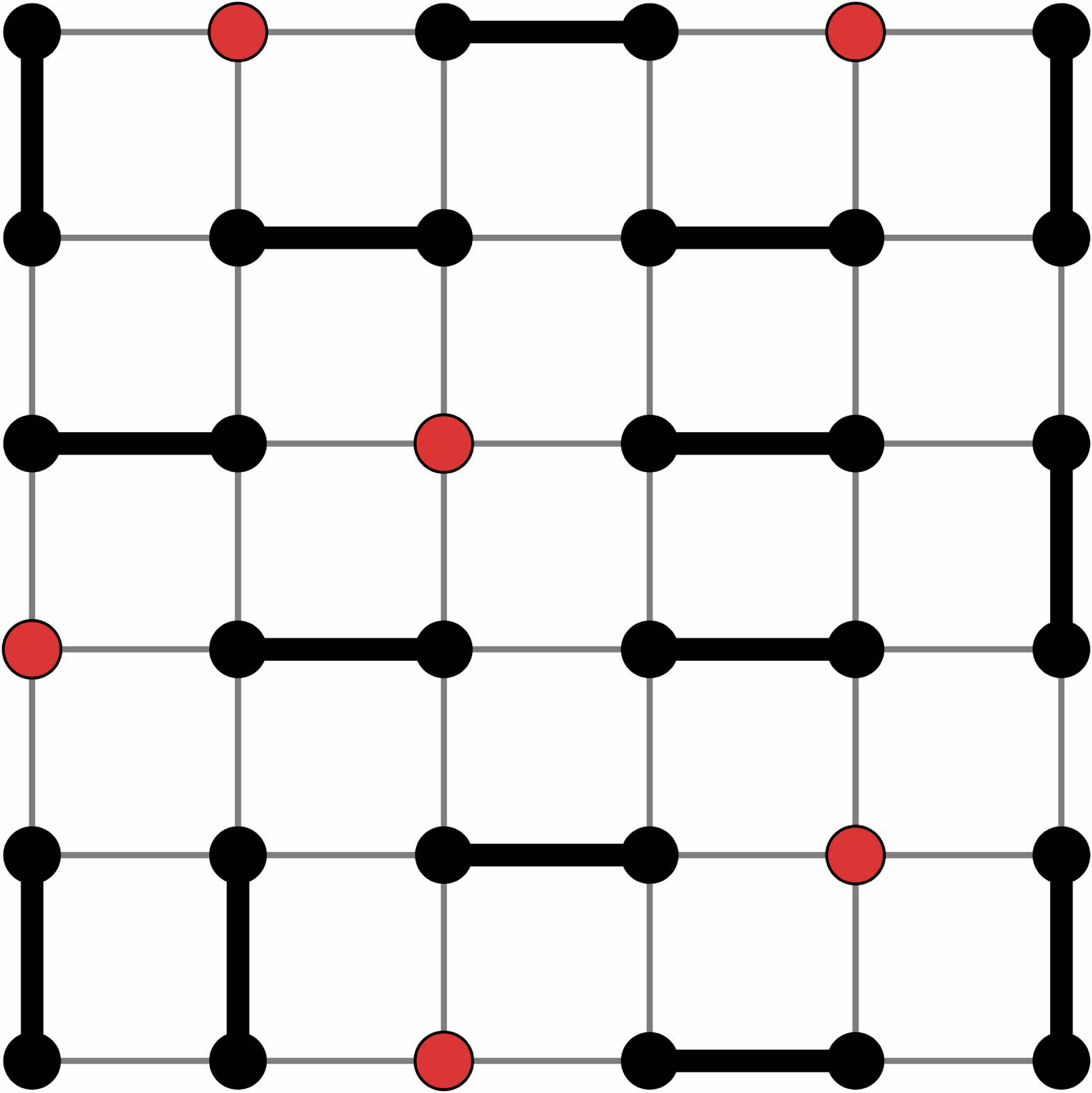}
\caption{(Color online) Typical dimer configuration for a square lattice of 
size $6\times 6$ without monomer (left) and with six monomers (right, red 
dots). 
}\label{fig_dimers}
\end{figure}%
For the general dimer problem where an arbitrary number of monomers are present 
-- the lattice sites that are not covered by the dimers are regarded as 
occupied 
by monomers -- there is no exact solution except in $\da$ where the solution 
can be expressed in terms of Chebyshev polynomials \cite{contuccistatistical}, 
on 
the complete graph and on locally tree-like graphs \cite{alberici2013solution}. 
\tc{We can also mention that the matrix transfer method was used to express the 
general monomer-dimer problem~\cite{lieb1967} (monomer density is not fixed), 
here the partition function, in terms of the maximum eigenvalue instead of a 
pfaffian. In particular a very efficient method based on variational corner 
transfer matrix has been found by Baxter~\cite{baxter1968dimers}, leading to
a precise approximation of thermodynamic quantities, such as the average 
dimer density which can be evaluated accurately as function of the dimer {\it 
activity}.}
For $\daaa$ lattices, no exact 
solution exists for the pure close-packed dimer problem. Recent advances concern 
the analytic solution of the problem where there is a single monomer on the 
boundary of a $\daa$ lattice 
\cite{2003dimers,wu06}, correlation functions for monomers 
located on the boundary \cite{izmailian05,priezzhev2008boundary} 
and localization phenomena of a monomer in the bulk \cite{bowick2007vacancy, 
poghosyan2011return}. 

The field of analytical solutions in the 
monomer-dimer model is still uncharted, but many rigorous results exist, e.g. 
location of the zeros of 
the partition function \cite{heilmann1970monomers,heilmann1972theory}, series 
expansions of the 
partition function  
\cite{nagle1966new} and exact recursion relation \cite{ahrens1981paving}. This lack of exact solution has been formalized in the context of computer science \cite{jerrum1987two}.
The importance of the dimer model in theoretical physics and combinatorics also
comes from the direct mapping between the square lattice Ising model 
without magnetic field and the dimer model on a decorated lattice 
\cite{mccoy1973two, 
kasteleyn1961statistics,fisher1961statistical,temperley1961dimer} \tc{and 
oppositely from the mapping of the   square lattice dimer model to a 
eight-vertex 
model \cite{baxter1972partition,wu1971ising} }.  Furthermore
the Ising model in a magnetic field can be mapped to the general 
monomer-dimer model \cite{heilmann1972theory}. 

Here we present a Grassmannian or fermionic formulation of the monomer-dimer 
model, which possesses an exact solution in terms of the product of two 
explicit pfaffians. We study the close-packed model, where an allowed dimer 
configuration has the property that 
each site of the lattice is paired with exactly one of its nearest neighbors, 
creating a dimer. In the simplest form, the number of dimers is the 
same in all the configurations, and the partition function is given by the 
equally-weighted average over all possible dimer configurations. In the 
following, we will include unequal fugacities, so that the average to be taken 
then includes nontrivial weighting factors.

\tc{ A early representation of the dimer model was introduced 
using 
Grassmann techniques~\cite{samuel1980use1,samuel1980use3}. A pair of these 
variables is attached to each site, 
preventing double occupancy of a site by two dimers. This leads to a direct 
representation of the partition function in terms of a fermionic
integral over a quartic action, from which diagrammatic 
expansions can be carried out.} 
We first review a very simple noncombinatorial interpretation of the 
$\daa$ dimer model based on the integration over Grassmann variables 
\cite{berezin,samuel1980use1,samuel1980use2}, and factorization principles for 
the density 
matrix \cite{hayn1994grassmann,plechko97}. A dimer model can be 
described with Boltzmann weights $t_x$ and $t_y$ of some coupling energy along 
the two directions. For example a magnetic field along one direction 
implies nonidentical weight values.
The partition function for a lattice of size $(L\times L)$ with $L$ even can
directly be written as
\bb\label{Q0def}
\mathcal{Q}_0=\int\prod_{m,n}\d\eta_{mn}
(1+t_x\eta_{mn}\eta_{m+1n})(1+t_y\eta_{mn}\eta_{mn+1})
\ee
where $\eta_{mn}$ are nilpotent and commuting variables satisfying 
\cite{palumbo97} $\eta_{mn}^2=0$, 
$\int \d\eta_{mn}\eta_{mn}=1$, and $\int \d\eta_{mn}=0$.
The integrals can be performed if we introduce, following closely Hayn and 
Plechko\cite{hayn1994grassmann}, a set of Grassmann variables 
$\{a_{mn},\bara{mn},b_{mn},\barb{mn}\}$ such that
\bb\nn
& &1+t_x\eta_{mn}\eta_{m+1n}=
\\ \nn
& &\int \d\bara{mn}\d a_{mn}e^{a_{mn}\bara{mn}}(1+a_{mn}\eta_{mn})
(1+t_x\bara{mn}\eta_{m+1n}),
\\ 
& &1+t_y\eta_{mn}\eta_{mn+1}=
\\ \nn
& &\int \d\barb{mn}\d b_{mn}e^{b_{mn}\barb{mn}}(1+b_{mn}\eta_{mn})
(1+t_y\barb{mn}\eta_{mn+1}).
\ee
This decomposition allows for an integration over the Grassmann variables $\eta_{mn}$, after 
rearranging the different link variables $A_{mn}=1+a_{mn}\eta_{mn}$, $\bar 
A_{m+1n}=1+t_x\bara{mn}\eta_{m+1n}$,
$B_{mn}=1+b_{mn}\eta_{mn}$ and $\bar B_{mn+1}=1+t_y\barb{mn}\eta_{mn+1}$. Then 
the partition function
becomes 
\bb
\mathcal{Q}_0=\Tr{a,\bar a,b,\bar b,\eta}\prod_{m,n}(A_{mn}\bar 
A_{m+1n})(B_{mn}\bar B_{mn+1}),
\ee
where we use the integration measure $\Tr{.}$ for the different Grassmannian 
and nilpotent variables with the adequate weights.

\begin{figure}
\includegraphics[scale=0.058,clip]{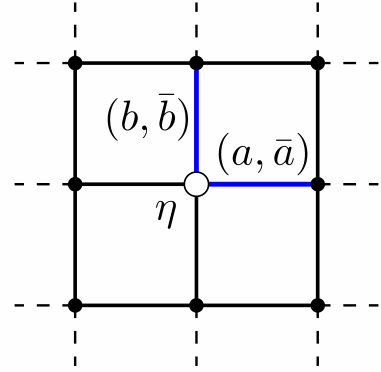}
\caption{(Color online) Variable configuration on site and links. At each site 
is associated a nilpotent variable $\eta$ such that $\eta^2=0$, and two pairs 
of Grassmann variables $(a,\bar a)$ and $(b,\bar b)$, one for each of the two 
directions.}
\end{figure}

The non-commuting link variables are then moved through the product in such a 
way that each $\eta_{mn}$ is isolated and can be integrated
directly. This rearrangement is possible in two dimensions thanks to the mirror 
symmetry introduced by Plechko \cite{plechko85a} for the $\daa$ Ising model. 
This also imposes the conditions $\bar 
A_{1n}=1$, $\bar A_{L+1n}=1$, $\bar B_{m1}=1$, and $\bar B_{mL+1}=1$, or 
$\bara{0n}=\bara{Ln}=\barb{m0}=\barb{mL}=0$ for open boundary conditions. One
finally obtains the following exact expression 
\bb\label{Q0ord}
\mathcal{Q}_0=\Tr{a,\bar a,b,\bar b,\eta}
\prodd{n}{}\Big (\prodg{m}{}\bar B_{mn}\prodd{m}{}\bar 
A_{mn}B_{mn}A_{mn}\Big ).
\ee
The integration over the $\eta_{mn}$ variables is performed recursively from 
$m=1$ to $m=L$ for each $n$. Each integration leads to a Grassmann 
quantity $L_{mn}=a_{mn}+b_{mn}+t_x\bara{m-1n}+(-1)^{m+1}t_y\barb{mn-1}$, which 
is moved to the left of the products over $m$ in \eref{Q0ord}, hence a minus 
sign is needed in front of each $\bar b$ crossed by $L_{mn}$ that is moved  
through the product of the $\bar B$ terms. 
Finally, the result $\mathcal{Q}_0=\Tr{a,\bar a,b,\bar 
b}\prod_{m,n}L_{mn}$ can be further rewritten by introducing additional 
Grassmann variables $c_{mn}$ such that $L_{mn}=\int \d c_{mn}\exp(c_{mn}L_{mn})$. This 
expresses $\mathcal{Q}_{0}$ as a Gaussian integral over variables $\{a,\bar 
a,b,\bar b,c\}$. The integration over variables $\{a,\bar a,b,\bar b\}$ can 
then be performed and, after anti-symmetrization of the expression, one obtains explicitly
\bw
\bb\label{Q0int}
{\mathcal Q}_0
=\int\prod_{m,n}\d c_{mn} \exp\sum_{m,n}\left [\ff 
t_x(c_{m+1n}c_{mn}-c_{m-1n}c_{mn})+\ff 
t_y(-1)^{m+1}(c_{mn+1}c_{mn}-c_{mn-1}c_{mn})
\right ]=\int\prod_{m,n}\d c_{mn} \exp{\mathcal{S}_0}.
\ee
\ew
Boundary conditions are now $c_{0,n}=c_{m,0}=c_{L+1,n}=c_{m,L+1}=0$. 
We consider a Fourier transformation 
satisfying open boundary conditions~\cite{hayn1994grassmann}, 
$c_{mn}=i^{m+n}\sum_{p,q=1}^Lc_{pq}f_{m}(p)f_{n}(q)$, 
where $f_{n}(p)=\sqrt{\frac{2}{L+1}}\sin\frac{\pi p n}{L+1}$ form an 
orthonormal set of functions $\sum_m f_m(p)f_m(q)=\delta_{pq}$. 
This leads to a block representation of the action in the momentum space, for momenta inside the 
reduced sector $1\le p,q\le L/2$. We note vectors 
$\vc{\alpha}=^t(c_{pq},c_{-pq},c_{p-q},c_{-p-q})$, where $-p$ is meant for 
$L+1-p$ and label $\alpha=\{p,q\}$. The four components of these vectors will 
be written $c_{\alpha}^{\mu}$ with $\mu=1\cdots 4$. Then 
$\mathcal{S}_0=\ffi c_{\alpha}^{\mu}M_{\alpha}^{\mu\nu}c_{\alpha}^{\nu}$, where 
the antisymmetric matrix $M$ is defined by
\[
M_{\alpha}=
\begin{pmatrix}
0 & 0 & -a_y(q) & -a_x(p)  \\
 0 & 0 & a_x(p) & -a_y(q) \\
a_y(q) & -a_x(p) & 0 & 0 \\
a_x(p) & a_y(q) & 0 & 0 \\
\end{pmatrix}
\]
with $a_x(p)=2t_x\cos\frac{\pi p}{L+1}$ and 
$a_y(q)=2t_y\cos\frac{\pi q}{L+1}$. The 
factor $i$ can be absorbed in a redefinition of the $c$'s  
variables. One simply obtains a product of cosine 
functions~\cite{hayn1994grassmann} as found by 
Kasteleyn, Temperley, and Fischer 
\cite{kasteleyn1961statistics,fisher1961statistical,temperley1961dimer}, since 
the pfaffian of 
$\prod_{\alpha}M_{\alpha}$ 
is the product $\prod_{p,q}[a_x(p)^2+a_y(q)^2]$ in the reduced sector of 
momenta, or
\bb
\mathcal{Q}_0=\prod_{p,q=1}^{L/2}\left [ 4t_x^2\cos^2\frac{\pi p}{L+1}+4t_y^2\cos^2\frac{\pi q}{L+1} \right ].
\ee
The matrix $M_{\alpha}$ is deeply related to the 
Kasteleyn~\cite{kasteleyn1961statistics} orientation matrix $K$ since 
$\mathcal{Q}_0=\Pf(\prod_{\alpha}M_{\alpha})=\Pf(K)$.

We consider now the case where an even number $n$ of monomers are present in the lattice at different fixed positions 
$\br_i=(m_i,n_i)$ with $i=1,\cdots,n$, see \efig{fig_dimers}. The partition 
function $\PQ{n}{\br_i}$, which we define as a correlation function between 
monomers after summing up over all dimer configurations, is the number of all 
possible dimer configurations with the 
constraint imposed by fixing the given monomer positions. 
This quantity is evaluated by inserting $\eta_{m_in_i}$ in
\eref{Q0def} at each monomer location, which prevents dimers 
from occupying these sites. It is useful to introduce additional 
Grassmann variables $h_i$ such that $\eta_{m_in_i}=\int \d 
h_i\exp(h_i\eta_{m_in_i})$. These insertions are performed at point $\br_i$ in 
\eref{Q0ord}, and the integration over $\eta_{m_in_i}$
modifies $L_{m_in_i}\rightarrow L_{m_in_i}+h_i$. However, by moving the 
anticommuting variables $\d h_i$ to the left of the remaining ordered 
product, a minus sign is introduced in front of each $\bar b_{mn_i-1}$ or $t_y$ 
coupling in $\bar B_{mn_i}$ for all $m>m_i$. We can replace more generally 
$\bar b_{mn-1}$ by $\epsilon_{mn}\bar b_{mn-1}$, such that $\epsilon_{mn_i}=-1$ 
for $m>m_i$, and $\epsilon_{mn}=1$ otherwise. The integration is then performed 
on the remaining $\{a,\bar a,b,\bar b\}$ variables as usual, so that 
$\PQ{n}{\br_i}$ 
can be expressed as a Gaussian form, with a sum of counter-terms corresponding 
to the monomer insertions, or
\bb\nn
\PQ{n}{\br_i}=\Tr{c,h}\e^{ 
\mathcal{S}_0+\sum_{ \{\br_i \}}c_{m_in_i}h_i+\mathcal{S}_I},
\\ \label{hole_action}
\mathcal{S}_I=2t_y\sum_{\{\br_i\}}\sum_{m=m_i+1}^L(-1)^{m+1}c_{mn_i-1}c_{
mn_i}.
\ee
%
\begin{figure}
\includegraphics[scale=0.12]{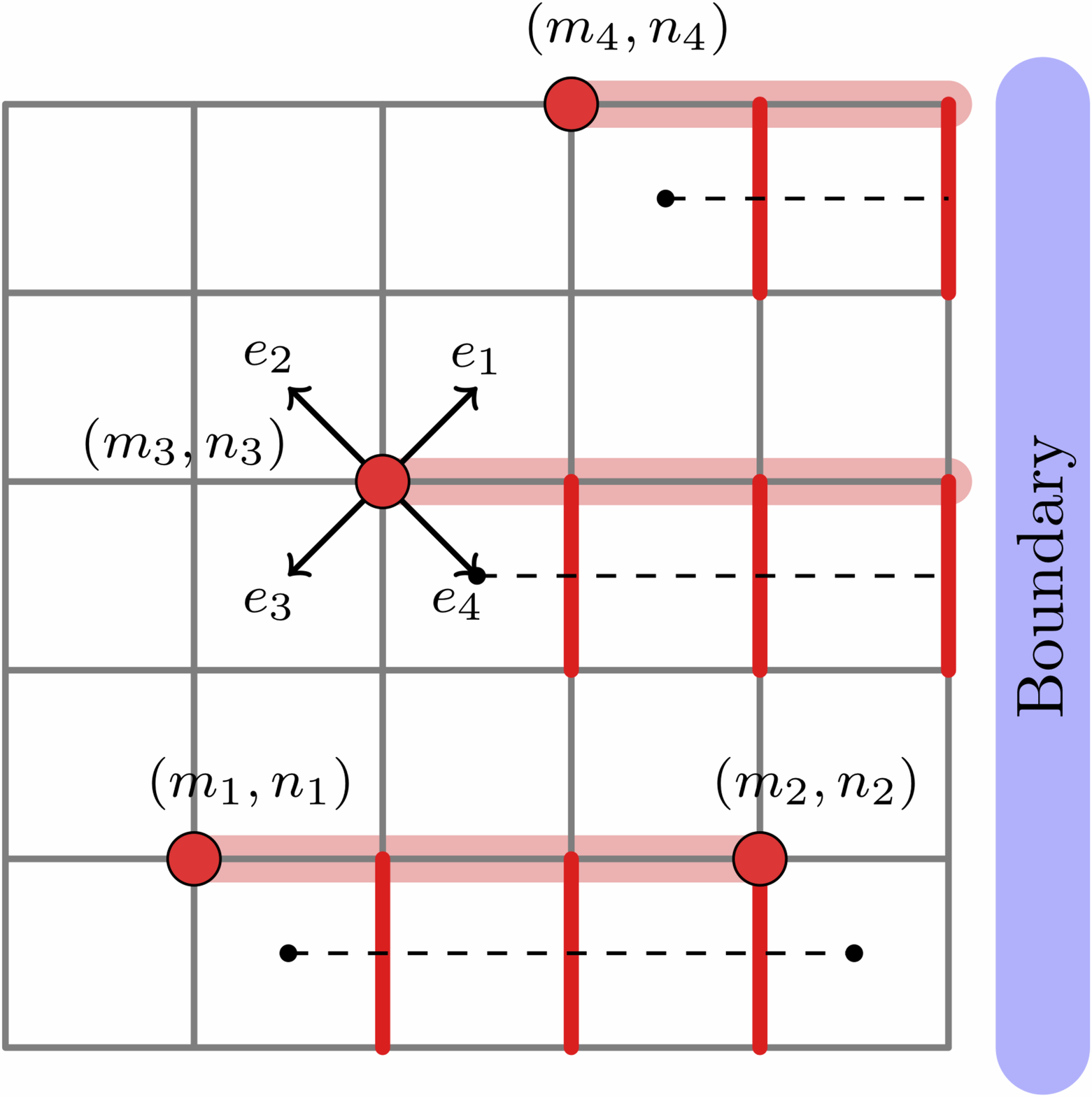}
\caption{(Color online) Typical configuration of the system with four monomers. 
The 
sign of the couplings $t_y$ are reversed (red links) along the black-dashed 
line \tc{(or disorder operator, see text)} that arises from moving the 
Grassmann fields conjugated to the defects 
toward the right boundary. Elementary vectors ${\bf{e}}_{i=1\cdots 4}$ are 
represented, and $\bf{e}_{4}$ indicates the starting location of the line of 
defects for the disorder operator.}\label{fig_2D}
\end{figure}
The contribution $\mathcal{S}_I$ corresponds to a line of defects, as shown in
\efig{fig_2D}. The addition of monomers is therefore equivalent to 
inserting a 
magnetic field $h_i$ at points $\br_i$, as well as a line of defect $c_{mn_i-1}c_{mn_i}$ running from the monomer 
position to the right boundary $m=L$. If two monomers have the same ordinate $n_i=n_j$, the line of defects will only
run between the two mononers and will not reach the boundary. 
\tc{This can be viewed as an operator acting on the links crossed by 
the line and running from a point on the dual lattice to the boundary 
on the right-hand side. More specifically, we can express the correlation 
functions, 
after integration over the fermionic magnetic fields $h_i$, as an average over 
composite fields}

\bw
\bb
\frac{\PQ{n}{\br_i}}{\mathcal{Q}_{0}}=\Big\<  \prod_{\{\br_{i}\}} c_{m_in_i} 
\exp \Big(2t_{y}\sum_{m=m_{i}+1}^{L}(-1)^{m+1}c_{mn_{i-1}}c_{mn_{i}} \Big)  
\Big\>_{0}=\Big\<  \prod_{\{\br_{i}\}} c_{m_in_i} \mu(\br_{i}+{\bf 
e_4})\Big\>_{0}=\Big\<  \prod_{\{\br_{i}\}} \Psi_4(\br_{i})\Big\>_{0},
\ee
\ew

\tc{where $\mu(\br+\bf{e}_i)$ is a disorder 
operator whose role is to change the sign of the vertical links across its 
path starting from vector $\br+\bf{e}_i$ on the dual lattice toward the right 
hand side, see \efig{fig_2D}. The integration $\<\cdots\>_0$ is performed 
relatively to the 
action $\mathcal{S}_0$. Elementary vectors $\bf{e}_i$ define a four-component 
fermionic field $\Psi_{\mu}(\br)=c_{mn} \mu(\br+{\bf 
e}_{\mu})$, which is the fermionic counterpart of the scalar field 
introduced for the Ising-spin 
model~\cite{kadanoff1971determination,polyakov1987gauge}. In the latter case, 
a linear differential equation can be simply found for
$\Psi_{\mu}(\br)=\sigma(\br)\mu(\br+{\bf e}_{\mu})$, with 
$\sigma(\br)=\pm 1$, leading to a Dirac equation. Here the general correlator 
between monomers is directly mapped onto the correlator between these fermionic 
composite fields.}
If we go back to \eref{hole_action}, the part of the field interaction can be 
Fourier transformed such that
$\sum_{ \{\br_i 
\}}c_{m_in_i}h_i=\sum_{p,q=1}^Lc_{pq}H_{pq}=\sum_{\alpha,\mu}c_{\alpha}^
{\mu}H_{\alpha}^{\mu}$.
The term $\mathcal{S}_I$ in the action can be written as $\ffi 
c_{\alpha}^{\mu}V_{\alpha\beta}^{\mu\nu}c_{\beta}^{\nu}$, 
with the perturbative matrix $V_{\alpha\beta}$ given by
\bw
\bb
V_{\alpha,\beta}=V_{pq,p'q'}=
\sum_{ \{\br_i \}  }2t_y(-1)^{n_i}
\left \{
\sum_{m=m_i+1}^L  f_{m}(p)f_{m}(p')\right \}
\Big(
f_{n_{i}-1}(q)f_{n_{i}}(q')-f_{n_{i}-1}(q')f_{n_{i}}(q)
\Big).
\ee
\ew
The different components $V_{\alpha\beta}^{\mu\nu}$ are given explicitly, for 
the first terms, by
$V_{\alpha\beta}^{11}=V_{pq,p'q'}$, $V_{\alpha\beta}^{12}=V_{pq,-p'q'}$,
$V_{\alpha\beta}^{21}=V_{-pq,p'q'}$, and so on. Then the full fermionic action
is $\mathcal{S}=\ffi c_{\alpha}^{\mu}W_{\alpha}^{\mu\nu}c_{\alpha}^{\nu}
+c_{\alpha}^{\mu}H_{\alpha}^{\mu}$ with antisymmetric
matrix $W_{\alpha\beta}^{\mu\nu}=\delta_{\alpha\beta}M_{\alpha}^{\mu\nu}
+V_{\alpha\beta}^{\mu\nu}$ 
satisfying $W_{\alpha\beta}^{\mu\nu}=-W_{\beta\alpha}^{\nu\mu}$.
By construction, this matrix can be represented as a block matrix of global 
size $(L^2\times L^2)$
\[
W=
\begin{pmatrix}
M_{\alpha=(1,1)} & V_{(1,1),(1,2)} & V_{(1,1),(1,3)} & \cdots &  \\
 -V_{(1,1),(1,2)} & M_{(1,2)} & V_{(1,2),(1,3)} & \cdots \\
-V_{(1,1),(1,3)} & -V_{(1,2),(1,3)} & M_{(1,3)} & \cdots  \\
\cdots & & & \\
\multicolumn{4}{c}{$\upbracefill$}\\
\multicolumn{4}{c}{\scriptstyle L^2/4 \;{\rm blocks}}\\
\noalign{\vspace{-\normalbaselineskip}}
\end{pmatrix},
\vspace{\normalbaselineskip}
\]
where each of the $(L^2/4)\times(L^2/4)$ blocks is a $(4\times4)$ matrix. Labels 
$\alpha$ are ordered with increasing momentum $(1,1),(1,2)\cdots 
(1,L/2),(2,1)\cdots$. Then $\PQ{n}{\br_i}$ can formally be 
written as $\PQ{n}{\br_i}=\Tr{c,h}\exp\left (\ffi 
c_{\alpha}^{\mu}W_{\alpha\beta}^{\mu\nu}c_{\beta}^{\nu}
+c_{\alpha}^{\mu}H_{\alpha}^{\mu}\right )$. The linear terms in 
$c_{\alpha}^{\mu}$ can be removed using a translation 
$c_{\alpha}^{\mu}\rightarrow 
c_{\alpha}^{\mu}+g_{\alpha}^{\mu}$, with
$g_{\alpha}^{\mu}=i(W^{-1})_{\alpha\beta}^{\mu\nu}H_{\beta}^{\nu}$. After a 
further rescaling of variables $c_{\alpha}\rightarrow 
i^{-1/2}c_{\alpha}$, one obtains
\bb\nn
\PQ{n}{\br_i}
=\Pf(W)
\Tr{h}
\exp\left [-\ffi(W^{-1})_{\alpha\beta}^{\mu\nu}H_{\alpha}^{\mu}H_{\beta}^{\nu}
\right ].
\ee
The fields $H_{\alpha}^{\mu}$ depend on $h_i$ through the identity 
$H_{\alpha}^{\mu}=\sum_{i=1}^{n}\Lambda_{i,\alpha}^{\mu}h_i$,
where coefficients $\Lambda_{i,\alpha}^{\mu}$ are expressed using a 
four-dimensional vector
${\bf \Lambda}_{i,\alpha}=f_{m_{i}}(p)f_{n_{i}}(q){\bf\Lambda}_i$.
The components of momentum-independent vector $\Lambda_i^{\mu}$ are
$(i^{m_i+n_i},-i^{-m_i+n_i},-i^{m_i-n_i},i^{-m_i-n_i})$. Its role is to fix 
whether the configuration of the monomers is allowed or not (in this case 
the correlator is zero). The final and compact expression for $\PQ{n}{\br_i}$ 
is 
then
\bb\label{corr2D}
\PQ{n}{\br_i}
=\Pf(W)\Pf(C),
\ee
where $C$ is a real $(n\times n)$ antisymmetric matrix with elements $C_{ij}=-i\Lambda_{i,\alpha}^{\mu}(W^{-1})_{\alpha\beta}^{\mu\nu}
\Lambda_{j,\beta}^{\nu}$.
The antisymmetry can be easily verified using the 
antisymmetry property of $W$ or $W^{-1}$.
\begin{figure}
\centering
\includegraphics[scale=0.66,clip]{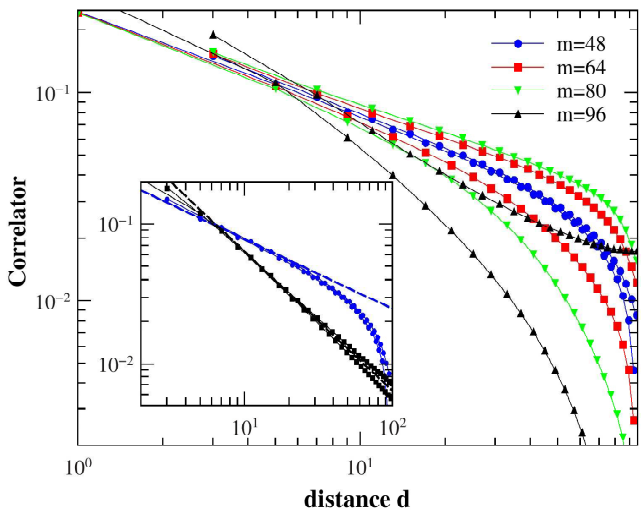}
\caption{(Color online) Correlation function 
$\PQ{2}{\br_1,\br_2}\mathcal{Q}_0^{-1}$ for two 
monomers on a lattice of size $L=96$ as function of their distance 
$\dst=|\br_2-\br_1|$. 
They are positioned vertically, at locations 
$\br_1=(m,L/2-k)$ 
and $\br_2=(m,L/2+k+1)$, with $\dst=2k+1$. The curves 
represent different abscissa $m$ successively from the right border ($m=L=96$) 
to the center 
of the lattice ($m=48$). Curves come by pair, with lower or higher 
correlations, depending if $k$ is even or odd. Inset: Correlation function 
using \eref{CijBound} for two monomers on the boundary (black square symbols), 
at locations $n_1=L/2-k$ and $n_2=L/2+k+1$, as function of 
their distance $\dst=2k+1$. Lattice size is $L=1000$. Asymptotic limit
$\frac{2}{\pi}\dst^{-1}$ 
(black dashed line) is shown for comparison. \tc{The 
bulk correlator (blue symbols $L=96$ and $m=48$) is also displayed, as well as 
its asymptotic limit $Bd^{-1/2}$ (blue dashed line). The value $B\simeq 
0.247$, see text.}}
\label{fig_2Dcorr}
\end{figure}%
$\mathcal{Q}_{n}$ is therefore a product of two pfaffians where the 
positions of the monomers are specified in both matrices $W$ and $C$. 
\tc{The factorization \eref{corr2D} can
generally be viewed as the product of a bulk and, by analogy, a {\it boundary} 
contribution. This can be found, for 
example, when a non-homogeneous magnetic field is applied at the surface of a 
2D Ising model~\cite{CF06}, by using Grassmann techniques as well. Here the 
term $\Pf(C)$ is due to the contribution of monomers in the bulk leading to a 
corrective factor in the free energy of order of the number $n$ of monomers, 
similar to a surface perturbation. Since the monomers are in the bulk, they 
contribute as well to the term $\Pf(W)$, which would otherwise, were the 
monomers located on the surface, be equal to $\mathcal{Q}_0$.} 
It is worth noting that a similar factorization was found
for the correlation function between two monomers in terms of the 
product of two spin-spin correlation functions of the Ising model at 
criticality~\cite{hartwig1966monomer,perk84}, due to the analogy of the dimer 
model with \tc{two Ising models (or a complex fermionic field theory), see 
Appendix for precise details}. 
It is, however, not obvious here to have such a direct identification with this 
result since the two pfaffians in \eref{corr2D} are of different nature. We can 
also mention that factorization of the correlation function exists in other 
models such as the one-dimensional XY chain~\cite{perk77}.
Matrix $V$ can be rewritten using additional matrices after considering the 
different components 
$(\mu,\nu)$. We can indeed express $V$ using four functions 
$u_k^{s=0,1}(\alpha,\beta)$, and 
$v_k^{s=0,1}(\alpha,\beta)$, for each monomer at location $\br_k=(m_k,n_k)$, 
with $m_k<L$, and such that $V_{\alpha\beta}=-2t_y
\sum_{\br_k}\sum_{s,s'=0,1}u_k^s(\alpha,\beta)\Gamma_{ss'}v_k^{s'}(\alpha,
\beta)$, with
\bb\nn
\Gamma_{01}&=&
\begin{pmatrix}
1 & 0 & 0 & 0  \\
 0 & 1 & 0 & 0 \\
0 & 0 & -1 & 0 \\
0 & 0 & 0 & -1 \\
\end{pmatrix},
\Gamma_{11}=
\begin{pmatrix}
0 & 1 & 0 & 0  \\
1 & 0 & 0 & 0 \\
0 & 0 & 0 & -1 \\
0 & 0 & -1 & 0 \\
\end{pmatrix},
\\ \nn
\Gamma_{00}&=&
\begin{pmatrix}
0 & 0 & 1 & 0  \\
 0 & 0 & 0 & 1 \\
-1 & 0 & 0 & 0 \\
0 & -1 & 0 & 0 \\
\end{pmatrix},
\Gamma_{10}=
\begin{pmatrix}
0 & 0 & 0 & 1  \\
0 & 0 & 1 & 0 \\
0 & -1 & 0 & 0 \\
-1 & 0 & 0 & 0 \\
\end{pmatrix}.
\ee
Functions $u$ and $v$ are specified by
\bb
&&u_k^s(\alpha,\beta)=\sum_{m=m_k+1}^L(-1)^{s(m+1)}f_{m}(p)f_{m}(p')
,\\
&&\frac{v_k^s(\alpha,\beta)}{(-1)^{sn_k}}=\big [
f_{n_{k}}(q)f_{n_{k}-1}(q')
+(-1)^s f_{n_{k}}(q')f_{n_{k}-1}(q)
\big ]\nn.
\ee
It is also worth noting that we have a similar structure in the real space, 
where the total action \eref{hole_action} is expressed by $\mathcal{S}=\ff 
c_{mn}W_{mn,m'n'}c_{m'n'}+\sum_{\br_i}c_{m_in_i}h_i$, with $W$
containing both the connectivity matrix $M$ and the contribution of the line 
of defects $V$. A direct computation also leads to the factorization
$\PQ{n}{\br_i}=\Pf(W)\Pf(C)$, where $C_{ij}=(W^{-1})_{m_in_i,m_jn_j}$ is a 
$(n\times n)$ antisymmetric matrix. 

Exact dimers enumeration algorithms \cite{krauth2006statistical} up to size of 
$10\times10$ has been widely used to compare with the theoretical prediction. 
For instance there are 636,072 different configurations of dimers with two 
monomers at
coordinates $\br_{1}=(2,3)$ and $\br_{2}=(7,5)$ on a $8\times8$ lattice, in 
accordance with the computation of $\mathcal{Q}_{2}(\br_{1},\br_{2})$ taking $t_{x}=t_{y}=1$. As possible other 
application, we could obtain the full partition function of the monomer-dimer 
model by summing up over all the possible number of monomers and over all the 
possible positions. The result for the $8\times8$ lattice is 
179,788,343,101,980,135~\cite{ahrens1981paving}, compared with the 12,988,816 
configurations without monomer.
In \efig{fig_2Dcorr}, we have solved numerically for a size $L=96$ the 
modified correlation function 
$\PQ{2}{\br_1,\br_2}\mathcal{Q}_0^{-1}=\Pf(M^{-1}W)\Pf(C)$, for two monomers 
at positions $\br_{1}=(m,L/2-k)$ and $\br_{2}=(m,L/2+k+1)$, $k=0\cdots 
L/2$, distant of $\dst=2k+1$. Due to finite-size effects, a curve for a given 
$m$ is distinguished depending on the parity 
of $k$. In the large size limit, this difference is indiscernible. \efig{fig_2Dcorr} shows the 
crossover between a behavior in $\dst^{-1}$ near the boundary ($m=96$) to a bulk 
behavior~\cite{fisher1963statistical} in $\dst^{-1/2}$ ($m=48$). \tc{The 
amplitude $B$ of the asymptotic two-point correlation function, which behaves 
like $B\dst^{-1/2}$, has been determined explicitly in the thermodynamic 
limit~\cite{perk84}, 
$B=2^{-3/4}A^{2}\approx 0.247$ with $A=2^{1/12}\e^{3\zeta'(-1)}$ and where 
$\zeta(s)$ is the Riemann zeta function. This value appears to be in good 
agreement with our numerical fit (see inset \efig{fig_2Dcorr}, dashed blue 
line).}
Interestingly, when the monomers are located exactly on the boundary 
($m=L$), $V=0$, and $W=M$, in this case $\PQ{n}{\br_i}=\mathcal{Q}_{0}\Pf(C)$, 
and it is straightforward to compute exactly the elements of matrix $C$. In the 
discrete case one obtains
\bw
\bb\label{CijBound}
C_{ij}=\frac{4\left [ (-1)^{n_i}-(-1)^{n_j} \right ]}{(L+1)^2}\sum_{p,q=1}^{L/2}
\frac{i^{1+n_i+n_j}t_y\cos\frac{\pi q}{L+1}\sin^2\frac{\pi p}{L+1}}
{t_x^2\cos^2\frac{\pi p}{L+1}+t_y^2\cos^2\frac{\pi q}{L+1}}
\sin\frac{\pi q n_i}{L+1}\sin\frac{\pi q n_j}{L+1}.
\ee
\ew
$C_{ij}$ are zero if $n_i$ and $n_j>n_i$ have the same parity. For example, 
fixing one monomer on the first site $n_1=1$ and taking $n_2=2k$, we have, 
for $t_x=t_y=1$ in the asymptotic limit $L\rightarrow\infty$ and large 
$k$, the following expansion $C_{12}\simeq 
\frac{2}{\pi}k^{-1}-\frac{3}{2\pi}k^{-5}$.
In the case $n_1=L/2-k$ and $n_2=L/2+k+1$, as shown in inset of 
\efig{fig_2Dcorr}, 
$C_{12}\simeq \frac{2}{\pi}\dst^{-1}-\frac{2}{\pi} 
\dst^{-3}$ instead, with $d=2k+1$ and amplitude $2/\pi$. This result is in 
agreement with the work of 
Priezzhev and Ruelle \cite{priezzhev2008boundary} on the scaling limit of the correlation functions 
of boundary monomers in a system of closely packed dimers in terms of a $\da$ 
chiral free fermion theory~\footnote{We can also 
mention that the result of 
the partition function of the dimer model with one monomer on the boundary 
\cite{2003dimers} can be easily recovered with our method.}. 

In summary, we presented a practical fermionic solution of the $\daa$ monomer-dimer
model on the square lattice, which allows for expressing the correlation
functions between monomers in terms of two pfaffians, and gave an explicit 
formula for boundary correlations. This can also be used for studying more 
general $n$-point correlation functions, thermodynamical 
quantities, or transport phenomena of monomers. Other lattice types, 
such as hexagonal and other boundary conditions, can be considered as well.

\tc{We are grateful to J. H. H. Perk for his knowledge in this domain 
and comments on the manuscript.}
This work was partly supported by the Coll\`ege Doctoral 
Leipzig-Nancy-Coventry-Lviv
(Statistical Physics of Complex Systems) of UFA-DFH.

\appendix*
\section{}\label{app}
In this section, we derive the continuum limit of the dimer action \eref{Q0int} 
and reformulate $\Sdimer$ in terms of two copies of Ising models. By an 
adequate change of variables~\cite{plechko97} $c_{mn}\rightarrow 
i^{3/2+m^{2}}c_{mn}$, the action $\mathcal{S}_0$ can 
be written as a complex fermion field theory:

\bw
\bb\nn
\mathcal{S}_0=
\sum_{m,n}\left [\ff t_x(c_{m+1n}c_{mn}-c_{m-1n}c_{mn})+\ffi 
t_y(c_{mn+1}c_{mn}-c_{mn-1}c_{mn})
\right ].
\ee
\ew

We can introduce the formal derivative using series expansions 
$c_{m+1n}=c_{mn}+\partial_x c_{mn}$ and $c_{mn+1}=c_{mn}+\partial_yc_{mn}$, 
up to first order in lattice elementary step, so that the action can be 
recognized as a purely kinetic form with no mass contribution:

\bb
\mathcal{S}_0=
\sum_{m,n}\left [ t_x\partial_x 
c_{mn}c_{mn}+it_y\partial_yc_{mn}c_{mn}
\right ].
\ee

It is convenient to define the following fields:

\bb\nn
c_{-}(m,n)&=&c_{2m2n},\;c_{+}(m,n)=c_{2m2n+1},
\\
\bar c_{-}(m,n)&=&c_{2m+12n+1},\;\bar c_{+}(m,n)=c_{2m+12n},
\ee

and express the previous action in terms of these fields only:

\bb\nn
\mathcal{S}_0&=&
-\sum_{m,n=1}^{L/2}\sum_{\sigma=\pm}\left [ t_x(c_{\sigma}\partial_x 
\bar c_{-\sigma}+\bar c_{\sigma}\partial_x 
c_{-\sigma})\right .
\\
&+&\left . it_y
(c_{\sigma}\partial_y c_{-\sigma}+\bar c_{\sigma}\partial_y 
\bar c_{-\sigma})
\right ].
\ee

Site variables $(m,n)$ now designate the locations of reduced cells containing 
four sites and take values between 0 and $L/2$. Field vectors $(c_{\sigma},\bar 
c_{\sigma})$ are composed of 
two independent components and describe two coupled Ising models 
labeled by index $\sigma=\pm$. This action can be diagonalized with a 
linear transformation, and new set of Grassmann variables:

\bb\nn
\varphi_{-}&=&\ff\left ( c_{-}+c_{+}+\bar c_{-}+\bar c_{+}\right ),
\\ \nn
\bar \varphi_{-}&=&\ff\left ( c_{-}+c_{+}-\bar c_{-}-\bar c_{+}\right 
),
\\ \nn
i\varphi_{+}&=&\ff\left ( c_{-}-c_{+}+\bar c_{-}-\bar c_{+}\right 
),
\\
i\bar \varphi_{+}&=&\ff\left ( c_{-}-c_{+}-\bar c_{-}+\bar c_{+}\right 
).
\ee

We obtain finally a diagonalized form for $\Sdimer$, defining the complex 
derivative in two-dimensions, $\partial=t_x\partial_x+it_y\partial_y$ and 
$\bar\partial=t_x\partial_x-it_y\partial_y$:

\bb\nn
\mathcal{S}_0=-
\sum_{m,n=0}^{L/2}\left ( 
\bar\varphi_{+}\bar\partial\bar\varphi_{+}
-\varphi_{+}\partial\varphi_{+}
-\bar\varphi_{-}\bar\partial\bar\varphi_{-}
+\varphi_{-}\partial\varphi_{-}
\right ).
\ee

Following Plechko~\cite{plechko97b}, it is useful to introduce Dirac 
matrices

\bb\nn
\sigma_{1}=
\begin{pmatrix}
0 & 1  \\
1 & 0 \\
\end{pmatrix},
\sigma_{2}=
\begin{pmatrix}
0 & -i  \\
i & 0 \\
\end{pmatrix},
\sigma_{3}=
\begin{pmatrix}
1 & 0  \\
0 & -1 \\
\end{pmatrix},
\ee

and define spinor $\Psi_{\sigma}=\begin{pmatrix} \varphi_{\sigma} \\
\bar\varphi_{\sigma}\end{pmatrix}$. It has to be noted that $\varphi_{\sigma}$ 
and $\bar \varphi_{\sigma}$ are not conjugated but independent Grassmann 
variables. The action can then be put into a compact expression,

\bb
\mathcal{S}_0=
\sum_{m,n=0}^{L/2}\sum_{\sigma=\pm}&^t\bar\Psi_{\sigma}\left (
\sigma_1\partial_1+\sigma_2\partial_2 \right )\Psi_{\sigma},
\ee

where $\bar \Psi_{\sigma}=i\sigma_{2}\Psi_{\sigma}$ and 
$\partial_1=t_x\partial_x$, $\partial_2=t_y\partial_y$. 
Here the resulting action is of Majorana form~\cite{plechko97b}, equivalent to 
two independent Ising models at criticality, since no mass term is present.

\bibliography{grassmann}

\end{document}